\documentclass[12pt]{article}
\newcommand{\ben}{\begin{enumerate}}
\newcommand{\een}{\end{enumerate}}
\newcommand{\be}{\begin{equation}}
\newcommand{\ee}{\end{equation}}
\newcommand{\bea}{\begin{eqnarray}}
\newcommand{\eea}{\end{eqnarray}}
\newcommand{\bc}{\begin{center}}
\newcommand{\ec}{\end{center}}

\newcommand{\A}{{\cal A}}
\renewcommand{\O}{\Omega}
\newcommand{\pa}{\partial}
\newcommand{\hpa}{\hat{\partial}}
\renewcommand{\d}{{\rm d}}
\renewcommand{\a}{{\bf a}}
\newcommand{\Ir}{Z\!\!\!Z}
\newcommand{\idty}{{\leavevmode{\rm 1\mkern -5.4mu I}}}
\newcommand{\Ibb}[1]{ {\rm I\ifmmode\mkern
            -3.6mu\else\kern -.2em\fi#1}}
\newcommand{\ibb}[1]{\leavevmode\hbox{\kern.3em\vrule
     height 1.2ex depth -.3ex width .2pt\kern-.3em\rm#1}}
\newcommand{\Cx}{{\ibb C}}
\newcommand{\Rl}{{\Ibb R}}
\newcommand{\Nl}{{\Ibb N}}

\begin{document}

\begin{tabbing}
\hspace*{11cm} \= GOET-TP 100/97  \\
               \> November 1997               \\
               \> 
\end{tabbing}

\begin{center}
       {\Large  \bf Deformations of Classical Geometries}
        \vskip.3cm
       {\Large \bf and Integrable Systems}
        \vskip1cm
        {\bf  Aristophanes DIMAKIS}
        \vskip.1cm
        {\it  Department of Mathematics, University of the Aegean}\\
        {\it  GR-83200 Karlovasi, Samos, Greece} \\
        e-mail: dimakis@aegean.gr
        \vskip.1cm
              and  
        \vskip.1cm
        {\bf  Folkert M\"{U}LLER-HOISSEN}
        \vskip.1cm
        {\it Institut f\"ur Theoretische Physik and MPI f\"ur Str\"omungsforschung} \\
        {\it Bunsenstrasse, D-37073 G\"ottingen, Germany} \\
        e-mail: fmuelle@gwdg.de
\end{center}

\begin{abstract}
A generalization of the notion of a (pseudo-) Riemannian space is proposed in a
framework of noncommutative geometry. In particular, there are parametrized families
of generalized Riemannian spaces which are deformations of classical geometries.
We also introduce harmonic maps on generalized Riemannian spaces into Hopf
algebras and make contact with integrable models in two dimensions.
\end{abstract}

\section{Introduction}

As a classical geometry we understand an $n$-dimensional  
Riemannian\footnote{Here and in the following `Riemannian' includes pseudo-Riemannian, 
i.e., the case of an indefinite metric.}
space $M$ which consists of a smooth orientable manifold $M$ and a metric 
tensor field 
\be
        g = g_{\mu \nu} \, \d x^\mu \otimes_\A \d x^\nu 
\ee
where $\A$ is the algebra of smooth function on $M$.
The metric induces a Hodge operator 
\be
      \star \; : \;   \Lambda^r(M) \rightarrow \Lambda^{n-r}(M)
\ee
where $\Lambda^r(M)$ is the space of differential $r$-forms on $M$. From the action of 
the Hodge operator we recover the (inverse) metric components with respect to the 
coordinates $x^\mu$ as follows,
\be
     g^{\mu \nu} = \star^{-1}( \d x^\mu \wedge \star \, \d x^\nu ) \; .
                            \label{metric_star}
\ee
\vskip.2cm

A generalization of classical geometries is obtained by generalizing the concept of
differential forms, accompanied with a suitable generalization of the Hodge operator.
The algebra of (ordinary) differential forms is then replaced by some `noncommutative'
differential algebra on $M$. Essentially, this means that we keep all the basic formulas
of the classical differential calculus but dispense with commutativity of functions and
differentials. A further generalization of geometries 
consists in replacing the underlying space $M$, or rather the (suitably restricted) algebra 
of functions on it, by some noncommutative associative algebra $\A$. All this will be made
more precise in section 2.
\vskip.2cm

Given a generalized Riemannian space, one can consider analogues of physical models
and dynamical systems on it. Of particular interest are generalized geometries and models 
which are deformations of classical geometries and models in the sense that they depend 
on some parameter in such a way that the basic algebraic relations become the classical ones 
when the parameter tends to a certain value. We then have the chance to study models 
which are `close' to known models of physical relevance. Section 3 is devoted to 
corresponding generalizations of harmonic maps into groups (or Hopf algebras), which 
are also known as (a class of) $\sigma$-models or principal chiral models. This is based 
on our previous work [1-5]. In subsection 3.3 we make an attempt to generalize the
latter to {\em non}commutative algebras. We have to stress, however, that this
is more a report on work in progress than something which has reached a satisfactory 
status. Section 4 contains some conclusions.

\section{Generalizations of classical geometries}
Let $\A$ be an associative algebra with unit element $\idty$. A {\em differential calculus} 
on $\A$ consists of a differential algebra $\O(\A)$ and an operator $\d$ which shares some
basic properties with the exterior derivative of the ordinary differential calculus on manifolds. 
A {\em differential algebra} is a $\Ir$-graded associative algebra (over $\Rl$, respectively $\Cx$)
\be
           \O (\A) = \bigoplus_{r \geq 0} \O^r (\A)
\ee 
where the spaces $\O^r (\A)$ are $\A$-bimodules and $\O^0(\A) = \A$. The operator $\d$
is a linear\footnote{Here and in the following {\em linear} means
linear over the respective field which is $\Rl$ or $\Cx$ in the cases under consideration.}
map  
\be
     \d \; : \quad \O^r (\A) \rightarrow \O^{r+1}(\A)
\ee
with the properties
\bea
       \d^2 &=& 0 \\
       \d (w \, w') &=& (\d w) \, w' + (-1)^r \, w \, \d w'
                           \label{Leibniz}
\eea
where $w \in \O^r(\A)$ and $w' \in \O (\A)$. The last relation is
known as the (generalized) {\em Leibniz rule}. We also require 
$\idty \, w = w \, \idty = w$ for all elements $w \in \O (\A)$. The 
identity $\idty \idty = \idty$ then implies $ \d \idty = 0 $. Furthermore,
it is assumed that $\d$ generates the spaces $\O^r(\A)$ 
for $r>0$ in the sense that $\O^r(\A) = \A \, \d \O^{r-1}(\A) \, \A$.

\subsection{Commutative algebras with noncommutative differential calculi and the 
           Hodge operator}
Let $\A$ be a {\em commutative} algebra, freely generated by elements $x^\mu$,
$\mu = 1, \ldots, n$. A differential calculus $(\O(\A), \d)$ is called $n$-{\em dimensional} 
if  \\
(1) $\d x^\mu$ is a left and also a right $\A$-module basis of $\O^1(\A)$,  \\
(2) $\O^r(\A) = \lbrace 0 \rbrace$ for $r > n$, but $\O^n(\A) \neq  \lbrace 0 \rbrace$,  \\
(3) $\mbox{dim} \, \O^r(\A) = \mbox{dim} \, \O^{n-r}(\A)$ as left as well as right
     $\A$-modules ($r = 0, \ldots, n$).
\vskip.2cm

In the following we consider a (freely generated) commutative algebra $\A$ with an 
$n$-dimensional differential calculus $(\O(\A), \d)$. A {\em generalized Hodge operator} 
is a linear invertible map
\be
      \star \, : \; \Omega^r \rightarrow \Omega^{n-r}  \qquad
                    r = 0, \ldots, n
\ee
such that\footnote{The `twist' in (\ref{star-covariance}) is dictated by certain
examples (which do not work with the alternative rule $\star \, (f  \, w) = f \, \star \, w$).
Note that the inverse of $\star$ satisfies $\star^{-1} ( f  \, w) = (\star^{-1} w) \, f $. }
\be              \label{star-covariance}
   \star \, (w \, f) = f \, \star \, w   \qquad \forall f \in \A, \, w \in \O^r(\A) \; .
\ee
Defining $\star$ on a basis of $r$-forms, this covariance property allows us to calculate
its action on any r-form. According to (\ref{metric_star}) we should be most
interested in the action of $\star$ on 1-forms and $n$-forms.
We call $(\A, \O(\A), \d, \star)$ an $n$-dimensional {\em generalized Riemannian 
space}. 
\vskip.2cm
\noindent
{\em Example.} Let $\A$ be the algebra of functions on the lattice $\Ir^n$ with
the $n$-dimensional differential calculus determined by
\be                       \label{latt_calc1}
     \lbrack \d x^{\mu} , x^{\nu} \rbrack = \delta^{\mu \nu} \, \d x^{\nu}
\ee
in terms of the canonical coordinates $x^{\mu}$ on $\Ir^n$ (cf [6]). As
a consequence, we have
\be     \label{d-discrete}
  \d f = \sum_{\mu = 1}^n [ f(x^1, \ldots, x^{\mu-1}, x^\mu +1, x^{\mu+1}, \ldots, x^n) 
                                          - f(x^1, \ldots, x^n) ] \, \d x^\mu
\ee
and
\be
     \d x^{\mu} \, \d x^{\nu} = - \d x^{\nu} \, \d x^{\mu}  \; .
\ee
This familiar anticommutativity of differentials does not extend to general 1-forms,
however, as in the ordinary calculus of differential forms. Let $\epsilon_{\mu_1
\ldots \mu_n}$ be totally antisymmetric with $\epsilon_{1 \ldots n} = 1$ and
$(\eta_{\mu \nu}) = \mbox{diag}(1, -1, \ldots, -1)$. We define
\be
 \star \, (\d x^{\mu_1} \cdots \d x^{\mu_r}) := {1 \over (n-r)!} \, \eta^{\mu_1 \nu_1}
 \cdots \eta^{\mu_r \nu_r} \, \epsilon_{\nu_1 \ldots \nu_r \kappa_1 \ldots \kappa_{n-r}} 
 \, \d x^{\kappa_1} \cdots \d x^{\kappa_{n-r}} \; .
\ee
What we have here is a discrete version of the $n$-dimensional Minkowski space.
Note that
\be
              \d x^\mu \star \d x^\nu = \star \, \eta^{\mu \nu}  \; .
\ee
In terms of the rescaled coordinates $x^{\mu'} := \ell^\mu \, x^\mu$ with constants 
$\ell^\mu >0$, (\ref{latt_calc1}) becomes
\be                      \label{latt_calc}
   \lbrack \d x^{\mu'} , x^{\nu'} \rbrack = \ell^\mu \, \delta^{\mu \nu} \, \d x^{\nu'} \; .
\ee
Ignoring the origin of the primed coordinates, this is a deformation of the algebraic
relations of the ordinary differential calculus on $\Rl^n$ (where differentials and functions
commute). For each coordinate a contraction $\ell^\mu \to 0$ can then be performed. 
In the new coordinates, the metric components are
\be
       g^{\mu' \nu'} := \star^{-1} ( \d x^{\mu'} \star \d x^{\nu'}) 
                             = \ell^\mu \, \ell^\nu \, \eta^{\mu \nu} 
\ee
and are thus witness to the rescaling.
A less trivial coordinate transformation is given by
\be
        y^\mu := (q^\mu)^{x^{\mu'} / \ell^\mu}
\ee
with $q^\mu \in \Cx \setminus \lbrace 0, 1 \rbrace$ and not a root of unity. This implies
\be
     \d y^\mu = {q^\mu -1 \over \ell^\mu} \, y^\mu \, \d x^{\mu'}  \, , \qquad
     \d x^{\mu'} \,  y^\mu  = q^\mu \,  y^\mu \, \d x^{\mu'}
\ee
and turns (\ref{latt_calc}) into the `quantum plane' relations
\be
    \d y^\mu \, y^\mu = q^\mu \, y^\mu \, \d y^\mu   
\ee 
(see also [7]). For different indices $\mu \neq \nu$, $\d y^\mu$ and $y^\nu$ simply 
commute. The components of the metric in the new coordinates $y^\mu$ are
\be
    g^{\mu \nu} := \star^{-1} ( \d y^{\mu} \star \d y^{\nu}) 
                        = (q^\mu -1) \,  (q^\nu -1) \, y^\mu \, y^\nu \, \eta^{\mu \nu}   \; .
\ee
One might expect that the inverse $g_{\mu \nu}$ defines an invariant object
via $g_{\mu \nu} \, \d y^\mu \otimes_\A \d y^\nu$, $\d y^\mu \otimes_\A 
g_{\mu \nu} \, \d y^\nu$ or $\d y^\mu \otimes_\A \d y^\nu \, g_{\mu \nu}$.
However, none of these expressions is equal to $\eta_{\mu \nu} \,  
\d x^\mu \otimes_\A \d x^\nu$, but differs by a factor which is a power of $q$.
{    }                                                               \hfill {\large  $\diamondsuit$}
\vskip.2cm

Using the Hodge operator, we define a scalar product on $\Omega^1$ 
by setting
\be               \label{scalar_prod}
          (\alpha,\beta) := \star^{-1}(\alpha \star \beta)   \; .
\ee
 From (\ref{star-covariance}) and the corresponding formula for the inverse of
$\star$ we obtain
\be 
        (\alpha , \beta \, f) = (\alpha \, f , \beta) \, ,  \qquad 
        (f \, \alpha , \beta) = (\alpha , \beta) \, f = f \, (\alpha , \beta) 
\ee
(since $\A$ is assumed to be commutative). 
The components of the scalar product are 
\be  
            g^{\mu \nu} := (\d x^\mu , \d x^\nu)  \; .
\ee
Let $y^\mu \in \A$. Then 
\be  
           dy^\mu = (\hat{\pa}_\nu y^\mu) \, \d x^\nu   
\ee
with generalized partial derivatives $\hat{\pa}_\nu$. We call $y^\mu$ `coordinates'
if $\hat{\pa}_\nu y^\mu$ is invertible. From the above properties 
of the scalar product we get
\be 
    g^{\mu' \nu'} = (\d y^\mu , \d y^\nu) = \hat{\pa}_\kappa y^\mu \, (\d x^\kappa , 
                            \hat{\pa}_\lambda y^\nu \, \d x^\lambda)  \; .                     
\ee
Let us now assume that the scalar product is {\em symmetric}, i.e., 
$(\alpha , \beta) = (\beta , \alpha)$ for all 1-forms $\alpha, \beta$, which means
\be
         \alpha \star \beta = \beta \star \alpha \; .    \label{star-symm}
\ee
In this case we have also
\be 
       (\alpha, f \, \beta) = f \, (\alpha , \beta) \; .  
\ee
and thus
\be 
     g^{\mu' \nu'} 
 =  \hat{\pa}_\kappa y^\mu \, \hat{\pa}_\lambda y^\nu \, 
     (\d x^\kappa , \d x^\lambda) 
 =  \hat{\pa}_\kappa y^\mu \, \hat{\pa}_\lambda y^\nu \, 
     g^{\kappa \lambda}   \; .  
\ee
To construct a tensor field from these components, the usual
tensor product $\otimes_{\cal A}$ is not the right one as long
as functions do not commute with differentials (see the
example above). In the case of a commutative algebra there is also
a tensor product, denoted as $\otimes_L$, which (besides
bilinearity over $\Rl$, respectively $\Cx$) satisfies 
\be 
 (f \, \alpha) \otimes_L (h \, \beta) = f \, h \, (\alpha \otimes_L \beta) \; . 
\ee
Then
\be  
    g := g_{\mu \nu}\, dx^\mu \otimes_L dx^\nu 
\ee
is a tensorial object.

\subsection{Noncommutative algebras and the Hodge operator}
The covariance property for the Hodge operator, as formulated in (\ref{star-covariance}), 
is not compatible with a {\em non}commutative algebra $\A$. A modification
is needed. Let ${}^\dag$ be an involution of $\A$. We generalize
the covariance rule as follows,\footnote{Note that this rule does not reduce to
our previous rule in the case of a commutative algebra $\A$ when the involution acts
nontrivially on $\A$.}
\be        
      \star \, (w \, f) = f^\dag \, \star w                \label{star_cov_dag}   
 \ee
so that 
\bea
   \star \, (w \, (f h) )=(f h)^\dag \star w = h^\dag \, f^\dag  \star w =
                                 h^\dag \star(w f) = \star \, ( (w f) \, h)   \; .
\eea
Again, we assume that the Hodge operator is an invertible map $\O^r(\A) \rightarrow
\O^{n-r}(\A)$ for some $n \in \Nl$. For its inverse (\ref{star_cov_dag}) implies
\be              
      \star^{-1}(f \, w) = (\star^{-1}w) \, f^\dag  \; . 
\ee
As a consequence, the scalar product on $\O^1(\A)$, defined again by 
(\ref{scalar_prod}), satisfies
\be
  (\alpha , \beta \, f) = (\alpha \, f^\dag, \beta) \, ,  \qquad
  (f \, \alpha , \beta) = (\alpha , \beta) \, f^\dag  \; .
\ee
Let us now assume that ${}^\dag$ extends to an involution of $\O(\A)$ so that
\be
                 (w \, w')^\dag = {w'}^\dag \, w^\dag   \; .
\ee
We still have to define how the exterior derivative $\d$ interacts with the involution.
Here we adopt the following rule 
\be 
      (\d w)^\dag = (-1)^{r+1} \, \d (w^\dag)          \label{d_dag}
\ee
for $w \in \O^r(\A)$ (cf [8], for example).
\vskip.2cm

We can now consistently impose the condition
\be                             
     (\star w)^\dag = \star^{-1}(w^\dag)      \label{star_dag}
\ee
since 
\be
   ( \star \, (w \, f) )^\dag = (f^\dag \, \star w)^\dag 
                                           = (\star \, w)^\dag \, f
                                           = [\star^{-1}(w^\dag)] \, f
                                           = \star^{-1}(f^\dag w^\dag)
                                           = \star^{-1}[(w f)^\dag]  \; .
\ee

\section{Generalized harmonic maps into matrix Hopf algebras}
Let $H$ be a matrix Hopf algebra (cf [9], in particular). This is a
Hopf algebra generated by elements $\a^i{}_j$, $i, j = 1, 
\ldots, N$. The coproduct $\Phi \, : \, \A \rightarrow \A \otimes \A$ is given by 
\be
      \Phi (\a^i{}_j) = \a^i{}_k \otimes \a^k{}_j
\ee
using the summation convention. The antipode $S$ satisfies 
\be
   S(\a^i{}_k) \, \a^k{}_j
      = \delta^i_j \, \idty 
      = \a^i{}_k \, S(\a^k{}_j)   \; .
\ee
In terms of the $N \times N$ matrix $\a = (\a^i{}_j)$ we have
$\Phi(\a) = \a \otimes \a$ and
\be  
      S(\a) \, \a = \left( \begin{array}{ccc} \idty & 0 & 0 \\ 
                                                            0 & \ddots & 0 \\
                                                            0 & 0 & \idty 
                              \end{array}  \right)
                   = \a \, S(\a)    
\ee                
in matrix notation. 
Let $(\A, \O(\A), \d, \star)$ be a generalized geometry in the sense of the preceeding section
and let us assume that the entries of $\a$ are constructed from elements of $\A$. The 
matrix of 1-forms
\be
               A := S(\a) \, \d \a
\ee 
then satisfies the identity
\be      \label{F=0}
          F := \d A + A A = 0  \; .
\ee
The field equation
\be
       \d \star A = 0      \label{hm_field_eq}
\ee
now defines a {\em generalized harmonic map} into a matrix Hopf algebra.\footnote{We may 
also call this a {\em generalized principal chiral model} or a {\em generalized $\sigma$-model}
(see [10], for example). Some related `noncommutative examples' can be found in [11].}
Note that we do not need the full Hodge operator here, but only its restriction to 1-forms, i.e.,  
$\star \, : \; \O^1(\A) \rightarrow \O^{n-1}(\A)$.
\vskip.2cm

A {\em generalized conserved current} of a generalized harmonic map is 
a 1-form $J$ which satisfies 
\be
                \d \star J = 0
\ee
as a consequence of the field equation (\ref{hm_field_eq}). 
We call a generalized harmonic map (completely) {\em integrable} if there is an infinite
set of independent\footnote{A convenient notion of independence in this context still has 
to be found. For example, in the case of classical models where integration is defined 
and the conserved currents lead to conserved charges, it may happen that a charge is 
a polynomial in some other charges. In an extreme case, we could get an infinite
tower of conserved charges as the set of polynomials in a single charge. We would
not like to talk about complete integrability in such a case.}
conserved currents.

\subsection{Integrable 2-dimensional generalized harmonic maps on
           commutative algebras}
 For 2-dimensional classical $\sigma$-models there is a construction of an
infinite tower of conserved currents [12]. This has been generalized
in [1-4] to harmonic maps on ordinary (topological) spaces, but with 
{\em non}commutative differential calculi, and values in a matrix group. 
In the following, we briefly recall the essential steps of our construction.
\vskip.2cm

Let us consider a generalized harmonic map on a  2-dimensional 
generalized Riemannian space (in the sense of subsection 2.1) which
satisfies the symmetry condition (\ref{star-symm}) and furthermore,
for $\alpha \in \O^1(\A)$,
\be              \label{closed-star-star} 
   \d \alpha = 0 \quad \Rightarrow \quad \alpha = \star \star \, \d \chi
\ee
with a function $\chi$. 
\vskip.2cm

Let us start with the $N \times N$ matrix 
\bea
    \chi^{(0)} := \left( \begin{array}{cccc}
                  \idty & 0 & \cdots & 0           \\
                   0    & \ddots & \ddots & \vdots \\
                 \vdots & \ddots & \ddots & 0      \\
                   0    & \cdots &   0    & \idty
                  \end{array} \right)   \; .
\eea
Then 
\be
    J^{(1)} := D \chi^{(0)} = (\d + A) \, \chi^{(0)} = A
\ee
is conserved as a consequence of the field equation. Using (\ref{closed-star-star})
this implies
\be
     J^{(1)} = \star \, \d \chi^{(1)}
\ee 
with an $N\times N$ matrix $\chi^{(1)}$. Now
\be
     J^{(2)} := D \chi^{(1)}  
\ee
is also conserved,
\be
     \d \star J^{(2)} = \d \star D \chi^{(1)} = D \star \d \chi^{(1)} 
   = D J^{(1)} = D^2 \chi^{(0)} = F  = 0  \; ,
\ee
since $ \d \star D  =  D \star \d  $ on $N \times N$ matrices with entries in $\A$. 
The latter follows from (\ref{star-covariance}), (\ref{hm_field_eq}) and (\ref{star-symm}) 
(cf [1,2]). Again, (\ref{closed-star-star}) implies
\be
    J^{(2)} := \star \, \d \chi^{(2)}  
\ee
with an $N \times N$ matrix $\chi^{(2)}$ of elements of $\A$. Now 
\be
    J^{(3)} := D \chi^{(2)}
\ee 
is another $N \times N$ matrix of conserved currents, and so forth. In this way we obtain 
an infinite set of (matrices of) conserved currents. There is no guarantee, however, 
that all these currents are really independent. It can happen, as in the case of the free 
linear wave equation on two-dimensional Minkowski space, that the higher conserved 
charges are just polynomials in a finite number of independent ones. 
\vskip.2cm

In this subsection we have considered a {\em commutative} algebra $\A$, but with a
{\em non}commu\-tative differential calculus. Even in this case a huge set of possibilities 
for integrable models arises and several examples have already been elaborated [1-4].
\vskip.2cm
\noindent
{\em Example.} We recall the following example from [1] (see also [2-5]).
Let $\A$ be the algebra of functions $f(t,x)$ on $\Rl \times \Ir$ which 
are smooth in the first argument. A differential calculus on $\A$ is then determined by
the relations
\be             \label{comm_rels}
   \lbrack \d t , t \rbrack = 0 \, , \quad 
   \lbrack \d x , x \rbrack = \ell \, \d x  \, , \quad
   \lbrack \d t , x \rbrack = \lbrack \d x , t \rbrack = 0   \; .
\ee
A Hodge operator,  restricted to 1-forms, is given by\footnote{Note that the
parameter $\ell$ does not appear in these relations. Hence they are not obtained by a
simple coordinate rescaling $x \mapsto x/\ell$ from the $\ell=1$ formulas. This has
to be distinguished from what we did in the example in subsection 2.1.}
\be
    \star \, \d t = - \d x \, , \qquad \star \, \d x = - \d t  \; .
\ee
The differential calculus and the Hodge operator satisfy the conditions 
(\ref{star-symm}) and (\ref{closed-star-star}). Let $\a = e^{- u}$ with
a function $u(t,x)$, so we consider only the case where $N=1$. Then, 
using (\ref{star-covariance}) and (\ref{d-discrete}), the 
field equation $\d \star A = 0$ turns out to be the equation of the nonlinear 
Toda lattice,
\be  
   \ddot{u}_k + {1 \over \ell^2} (e^{u_k-u_{k+1}} 
                      - e^{u_{k-1} - u_k}) = 0 
\ee
where $u_k(t) := u(t, k \, \ell)$. In the limit as $\ell \to 0$ the generalized
geometry tends to that of the 2-dimensional Minkowski space and the above
field equation becomes the linear wave equation.
We refer to the references mentioned above for details and also 
for matrix generalizations of the Toda lattice (i.e., $N > 1$).
{    }                                                               \hfill {\large  $\diamondsuit$}

\vskip.2cm
\noindent
A new example is presented in the following subsection.

\subsection{Another example}
In [6] we found in particular the following differential calculus,
\bea 
 [\d t , t] = b \, \d t \, , \quad  [\d t , x] = b \, \d x \, , 
\quad [ \d x , t] = b \, \d x \, ,  \quad  [\d x , x] = - 
{a^2 \over b} \, \d t       \label{str_calc}
\eea
with real constants $a,b \neq 0$.
In terms of the complex variable $ z = t/b + i \, x/a $, the above
commutation relations read
\bea 
 [ \d z , z ] = 2 \, \d z \, , \quad  [ \d z , \bar{z}] = 0 \, , \quad
 [ \d \bar{z}, z ] = 0 \, , \quad [ \d \bar{z}, \bar{z}] = 2 \, \d \bar{z}
\eea
where $\bar{z}$ is the complex conjugate of $z$. In the complex
coordinates $z, \bar{z}$ we thus have a two-dimensional lattice 
differential calculus (cf the example in subsection 2.1). The relations 
(\ref{str_calc}) extend to arbitrary functions $f_t(x) := f(t,x)$ as follows,
\bea
  \d t \, f_t & = &  C_x f_{t+b} \, \d t + {b \over a} \, S_x f_{t+b} \, \d x  
                        \nonumber  \\
  \d x \, f_t & = & -{a \over b} \, S_x f_{t+b} \, \d t + C_x f_{t+b} \, \d x 
                        \nonumber  \\
  f_t \, \d t & = &  \d t \, C_x f_{t-b} - {b \over a} \, \d x \, S_x f_{t-b} 
                        \nonumber  \\
  f_t \, \d x & = & {a \over b} \, \d t \, S_x f_{t-b} + \d x \, C_x f_{t-b} 
\eea
where the operators $C_x, S_x$ are defined by
\bea 
    (C_x f)(x) & := & {1\over 2} \, [ f(x+ia) + f(x-ia)]  \\
    (S_x f)(x) & := & {1\over 2i} \, [ f(x+ia) - f(x-ia) ]  
\eea
acting on a function of $x$. They satisfy
\bea
    C_x (f h) & = & (C_x f)(C_x h) - (S_x f)(S_x h)  \\
    S_x (f h) & = & (S_x f)(C_x h) + (C_x f) (S_x h) 
\eea
and
\bea
          C_x^2 f + S_x^2 f = f  \; . 
\eea
 Furthermore, we have
\bea
     \d f_t = {1 \over b} \, (C_x f_{t+b}- f_t) \, \d t + 
                 {1 \over a} \, (S_x f_{t+b}) \, \d x \; .
\eea
Using $ \d t \, \d t = \d x \, \d x = \d t \, \d x + \d x \, \d t = 0$, 
which follows from (\ref{str_calc}) by application of the exterior derivative $\d$,
a Hodge operator which satisfies (\ref{star-symm}) is given by
\bea  
        \star \, \d t = \sigma \, \d t + {b \over a} \, \kappa \, \d x \, , \qquad
        \star \, \d x = {a \over b} \, \kappa \, \d t - \sigma \, \d x  
\eea
with constants $\kappa, \sigma$. When $\kappa^2+\sigma^2=1$, it has the 
property $\star \star \, w = w$ for all 1-forms $w$. Together with the
property of the differential calculus that every closed 1-form is 
exact\footnote{The proof is simple. It is a slight variation of the proof of a 
Lemma in [4].}, this implies that (\ref{closed-star-star}) holds. Furthermore, 
a direct calculation shows that also (\ref{star-symm}) is satisfied. 
The above constraint for the constants $\kappa, \sigma$ is solved by writing
$\sigma=\sin \theta$ and $ \kappa=\cos \theta$ with a parameter $\theta$.
\vskip.2cm

With $\a = e^{-u_t}$ we obtain
\bea
  A = e^{u_t}\, \d e^{-u_t} 
  & = & {1 \over b} \, (e^{u_t}C_x e^{-u_{t+b}}-1) \, \d t
             +{1\over a} \, (e^{u_t}S_x e^{-u_{t+b}}) \, \d x  \nonumber  \\
  & = & \d t \, {1 \over b} \, (e^{-u_t}C_x e^{u_{t-b}}-1) - 
             \d x \, {1 \over a} \, (e^{-u_t}S_x e^{u_{t-b}} )    \; .
\eea
Application of the Hodge operator leads to
\bea
           \star A 
  &=& -{1\over b} \, [\sin \theta + e^{-u_t} \, \sin(a \, \pa_x-\theta) 
           \, e^{u_{t-b}}] \, \d t                                \nonumber \\
  & & + {1 \over a} \, [e^{-u_t} \, \cos(a \, \pa_x-\theta) \, e^{u_{t-b}}
                                   - \cos \theta] \, \d x 
\eea
and the field equation $\d \star A = 0$ takes the form
\bea
   e^{u_t} \, \cos (a \, \pa_x-\theta) \, e^{-u_{t+b}} = 
    e^{-u_t} \, \cos (a \, \pa_x-\theta) \, e^{u_{t-b}} 
\eea  
which, admittedly, is a rather unfamiliar equation.

\subsection{Generalization to noncommutative algebras}
In order to generalize the construction of conservation laws to {\em non}commutative
algebras $\A$, we impose some conditions in addition to those already introduced in 
subsection 2.2. In particular, we make the assumption that  for each 
$r = 0, \ldots, n$ there is a constant $\epsilon_r \neq 0$ such that\footnote{An apparently
weaker assumption would be: for each $w \in \O^r(\A)$ exists a constant 
$\epsilon_w$ such that $\star \star w = \epsilon_w \, w$. However, this reduces to
our previous assumption as follows. Since $\star \star$ is linear, we have
$\epsilon_{w+w'} \, (w+w') = \star \star (w+w') = \star \star w + \star \star w' 
=  \epsilon_{w} w + \epsilon_{w'} w'$ and thus
$ (\epsilon_{w+w'}-\epsilon_w) w = (\epsilon_{w+w'}-\epsilon_{w'}) w'$
for arbitrary $w,w' \in \O(\A)$. If $w, w'$ are linearly independent, then
$\epsilon_w = \epsilon_{w'}= \epsilon_{w+w'}$. Furthermore, 
$c \, \epsilon_w w = c \, \star \star w = \star \star (c \, w) = \epsilon_{c w} \, c \, w$
implies $\epsilon_{c w} = \epsilon_w$ where $c \in \Cx$. It follows that $\epsilon$ 
is constant on $\O^r(\A)$, i.e., $ \epsilon_w = \epsilon_r \; \forall \, w \in \O^r(\A)$.} 
\be             
        \star \star w = \epsilon_r \, w     \qquad     \forall w \in \O^r   
\ee
respectively,
\be 
    \star \, w = \epsilon_r^\dag \, \star^{-1}w   \; . \label{star_eps}
\ee
Applying the involution and using (\ref{star_dag}) we find
\be  
        {1 \over \epsilon_r^\dag} \, \star (w^\dag) 
     = \star^{-1} (w^\dag)  
     = (\star w)^\dag 
     =  \epsilon_r \, (\star^{-1} w)^\dag 
     =  \epsilon_r  \, \star (w^\dag)
\ee
and thus
\be  
        \epsilon_r^\dag = { 1 \over \epsilon_r }   \; .
\ee
Instead of the symmetry condition (\ref{star-symm}) we impose the condition 
\be                  \label{dag_star_eps}
     (\alpha \star \beta)^\dag = \epsilon_n^\dag \, \beta \star \alpha                                                
\ee
where $\alpha, \beta \in \O^1(\A)$. This is consistent with (\ref{star_cov_dag}) since
\be
      [ \alpha \star (\beta \, f) ]^\dag
   = [ \alpha \, f^\dag \, \star \beta]^\dag 
   = \epsilon_n^\dag \, \beta \star (\alpha \, f^\dag)
   = \epsilon_n^\dag \, (\beta \, f) \star \alpha          \; .
\ee 
As a consequence, we find
\bea
       (\alpha , \beta)^\dag &=& [ \star^{-1}(\alpha \star \beta) ]^\dag 
   = [ \star^{-1} \, \epsilon_n \, (\beta \star \alpha)^\dag ]^\dag
   = \epsilon_n \, \star (\beta \star \alpha)
   = \epsilon_n \, \epsilon_n^\dag \, (\beta, \alpha)  \nonumber \\
  &=& (\beta , \alpha)   \; .
\eea

A crucial step in the construction of conserved currents for harmonic
maps on commutative algebras in subsection 3.1 is the identity 
$\d \star D = D \star \d$. A suitable generalization is now obtained as 
follows. First, we have
\bea
       ( \d \star \d \chi^i{}_j )^\dag 
  =  \d (\star \, \d \chi^i{}_j)^\dag 
  =  \d \star^{-1} (\d \chi^i{}_j)^\dag
  =  - \d \star^{-1} \d (\chi^i{}_j)^\dag
  = - \epsilon_1 \, \d \star \d (\chi^i{}_j)^\dag
\eea
using (\ref{d_dag}), (\ref{star_dag}), again (\ref{d_dag}) and then (\ref{star_eps}). 
 Furthermore,
\bea
       [ \d (\chi^k{}_j)^\dag \star A^i{}_k ]^\dag  
    = \epsilon_n^\dag \, A^i{}_k \star \d (\chi^k{}_j)^\dag  
\eea 
using (\ref{dag_star_eps}). Hence
\bea
    \d \star D \chi^i{}_j & = & \d \star (\d \chi^i{}_j + A^i{}_k \, \chi^k{}_j) 
    = \d \star \d \chi^i{}_j + \d ((\chi^k{}_j)^\dag \star A^i{}_k)   \nonumber \\
 & = & \d \star \d \chi^i{}_j + \d (\chi^k{}_j)^\dag \star A^i{}_k +
           (\chi^k{}_j)^\dag \d \star A^i{}_k   \nonumber  \\
 & = & [ (\d \star \d \chi^i{}_j)^\dag + (\d (\chi^k{}_j)^\dag \star A^i{}_k)^\dag ]^\dag 
                                                                 \nonumber \\
 & = & [- \epsilon_1 \,	\d \star \d(\chi^i{}_j)^\dag + \epsilon_n^\dag \, A^i{}_k \star
            (\chi^k{}_j)^\dag ]^\dag
\eea
using $\d \star A = 0$. Consequently, if $\epsilon_1 = - \epsilon_n^\dag$ we have
\be
   \d \star D (\chi^\dag) = - (\epsilon_1^\dag \, D \star \d \chi)^\dag   \; .
\ee
 For $n=2$ and assuming (\ref{closed-star-star}), the construction of conservation laws now
generalizes to the case of a {\em non}commutative algebra $\A$. 
\vskip.2cm
\noindent
{\em Example.}
Let $\A$ be the Heisenberg algebra with the two generators $q$ and $p$ satisfying
\be   
       [q,p] = i \, \hbar  \; .
\ee
In the simplest differential calculus on $\A$ we have
\be
     [\d q , f] =0 \, ,  \qquad  [\d p , f] = 0   \label{Hbg_dqp_f}
\ee
for all $f \in \A$ (see also [13]). It follows that
\be
         \d f  = (\hpa_q f) \, \d q + (\hpa_p f) \, \d p
\ee
where the generalized partial derivatives are given by
\be
         \hpa_q f := -{1\over i\hbar}[p,f] \, ,   \qquad
         \hpa_p f := {1\over i \hbar}[q,f]   \; .
\ee
Moreover, the relations (\ref{Hbg_dqp_f}) imply
\be   
   \d q \, \d q = 0 \, ,  \quad
   \d q \, \d p + \d p \, \d q =0 \, , \quad \d p \, \d p = 0  \; .
\ee
As an involution we choose hermitean conjugation with
$q^\dag = q, \, p^\dag = p$. 
A $\star$-operator satisfying the conditions (\ref{star_eps}) and (\ref{dag_star_eps}) 
is determined by
\be
   \star \, 1 = \d q \, \d p \, , \quad 
   \star \, \d q = \d p  \, ,  \quad 
   \star \, \d p = - \d q \, , \quad
   \star \, (\d q \, \d p) = 1   
\ee
(so that $\epsilon_0 = \epsilon_2 =1, \, \epsilon_1 = -1$). Now we consider a 
generalized harmonic map with values in the group of unitary elements $U$ of $\A$ 
which satisfy $U^\dag U= \idty =U U^\dag$. With
\bea
      A &:=& U^\dag \, \d U    \nonumber \\
          &=& - {1\over i \hbar} \, (U^\dag p \, U-p) \, \d q 
                   + {1 \over i \hbar} \, (U^\dag q \, U-q) \, \d p
\eea
and thus
\be
  \star \, A = {1\over i \hbar} \, (U^\dag p \, U-p) \, \d p 
                 + {1\over i \hbar} \, (U^\dag q \, U-q) \, \d q   
\ee
the field equation $\d \star A = 0$ becomes
\be
     [p , U^\dag p \, U] + [q , U^\dag q \, U] = 0   \; .
\ee
In terms of $P := U^\dag p \, U , \, Q := U^\dag q \, U$ this takes the form
\be  
  [p,P] + [q,Q] = - i \, \hbar \, (\hpa_q P - \hpa_p Q) = 0 \; .  
\ee
If $\d \alpha = 0$ for a 1-form $\alpha$ implies $\alpha = \d F$ with some $F \in \A$,
then also (\ref{closed-star-star}) holds and all the required conditions are fulfilled.  
On the level of formal power series in $q$ and $p$, every closed 1-form is indeed exact.
Of course, one would like to substantiate these results on the level of functional
analysis which, however, is beyond the scope of this work.
{    }                                                               \hfill {\large  $\diamondsuit$}

\section{Conclusions}
We proposed a generalization of Riemannian geometry and harmonic maps
within a wide framework of noncommutative geometry. It centers around a generalization
of the Hodge operator to (noncommutative) differential calculi on associative algebras.  
The main motivation originated from previous work where we recovered lattice gauge theory
from a noncommutative geometry on $\Rl^n$ [6] and where we realized that, in particular,
the nonlinear Toda lattice model can be expressed as an integrable harmonic map equation, 
but with respect to a simple deformation of the ordinary differential calculus  [1-4] (cf the 
example in subsection 3.1). 
\vskip.2cm

In the present work we also addressed the case of  `noncommutative spaces'. Further work 
is certainly needed to understand the significance of  `noncommutative integrable models'
and more examples have to be elaborated. It should also be possible to weaken the restrictions 
some more which we imposed above in order to generalize the classical construction of
conservation laws for 2-dimensional $\sigma$-models to noncommutative spaces. 
The new material which we presented in subsection 3.3 has a fair chance to 
guide us to a class of interesting models in the same way as the intermediate step to
noncommutative differential calculi on ordinary spaces led us  [1] to physical models like
the nonlinear Toda chain.

\section*{Acknowledgments}
 F. M.-H. is grateful to Metin Arik and Rufat Mir Kasimov for the kind invitation to 
present the material of this paper at the conference on {\it Quantum Groups, Deformations 
and Contractions}.

\end{document}